\documentclass[aps,prd,onecolumn,groupedaddress,nofootinbib,preprintnumbers,superscriptaddress]{revtex4}  
\usepackage{graphicx,mathtools,tabu}
\usepackage{epstopdf}
\usepackage{amsmath}
\usepackage{amsfonts}
\usepackage[colorlinks=true,citecolor=red,linkcolor=blue,breaklinks=true]{hyperref}
\usepackage{accents}
\usepackage[figure, table]{hypcap}
\usepackage{amssymb}
\usepackage{appendix}
\usepackage{comment}
\usepackage{bbold}
\usepackage{color}
\usepackage{slashed}
\usepackage{subfigure}
\usepackage{setspace}
\usepackage{footnote}
\usepackage{multirow}
\usepackage{braket}
\usepackage[normalem]{ulem}
\usepackage[utf8]{inputenc}
\usepackage{mathrsfs}
\usepackage{longtable}
\usepackage{enumitem}
\usepackage{booktabs} 

\LTcapwidth=\textwidth

\usepackage{url}
\usepackage{wrapfig}
\usepackage{braket}

 \def\be   {\begin{equation}}  
 \def\ee   {\end{equation}}
 \def\ba   {\begin{array}}     
  \def\ea   {\end{array}}
 \def\bea  {\begin{eqnarray}}  
  \def\eea  {\end{eqnarray}}
 \def\bean {\begin{eqnarray*}}  
 \def\eean {\end{eqnarray*}}
 \def\ga   {\gamma}

  \def\al   {\alpha}
  \def\be {\beta}
    
      \def\r {\rho}

\def \R  {R_{\rm CCSN}}

\def\to {\rightarrow}

\newcommand{\g}{{\rm\ g}}

\newcommand{\MeV}{{\rm\ MeV}}

\begin{document}

\preprint{IPPP/22/29}
\title{The diffuse supernova neutrino background as a probe of late-time neutrino mass generation} 

\author{Andr\'{e} de Gouv\^{e}a}
\email{degouvea@northwestern.edu}
\affiliation{Northwestern University, Department of Physics \& Astronomy, 2145 Sheridan Road, Evanston, IL 60208, USA}
\author{Ivan Martinez-Soler}
\email{imartinezsoler@fas.harvard.edu}
\affiliation{Department of Physics \& Laboratory for Particle Physics and Cosmology, Harvard University, Cambridge, MA 02138, USA}
\author{Yuber F. Perez-Gonzalez}
\email{yuber.f.perez-gonzalez@durham.ac.uk}
\affiliation{Institute for Particle Physics Phenomenology, Durham University, South Road DH13EL, Durham, United Kingdom}
\author{Manibrata Sen}
\email{manibrata@mpi-hd.mpg.de}
\affiliation{Max-Planck-Institut f\"ur Kernphysik, Saupfercheckweg 1, 69117 Heidelberg, Germany}

\begin{abstract}
    The relic neutrinos from old supernova explosions are among the most ancient neutrino fluxes within experimental reach.
    Thus, the diffuse supernova neutrino background (DSNB) could teach us if neutrino masses were different in the past (redshifts $z\lesssim 5$).
    Oscillations inside the supernova depend strongly on the neutrino mass-squared differences and the values of the mixing angles, rendering the DSNB energy spectrum sensitive to variations of these parameters.
    Considering a purely phenomenological parameterization of the neutrino masses as a function of redshift, we compute the expected local DSNB spectrum here on Earth.
    Given the current knowledge of neutrino oscillation parameters, specially the fact that $|U_{e3}|^2$ is small, we find that the $\nu_e$ spectrum could be significantly different from standard expectations if neutrinos were effectively massless at $z\gtrsim1$ as long as the neutrino mass ordering is normal.
    On the other hand, the $\overline{\nu}_e$ flux is not expected to be significantly impacted. Hence, a measurement of both the neutrino and antineutrino components of the DSNB should allow one to test the possibility of recent neutrino mass generation.
\end{abstract}

\maketitle

\section{Introduction}
\setcounter{equation}{0}
The discovery of neutrino oscillations in the last century established without a doubt that neutrinos are massive. Neutrino oscillations provide precise information on the neutrino mass-squared differences but are independent from the absolute masses of the neutrinos. 
Data from neutrino oscillation experiments can be used to constrain the sum of the neutrino masses to $\sum m_i \gtrsim 0.058\,{\rm eV}$, $i=1,2,3$ in case of the Normal Mass Ordering (NO) $(m_3>m_2>m_1)$ or $\sum m_\nu \gtrsim 0.1\,{\rm eV}$ in the case of the Inverted Mass Ordering (IO) $(m_2>m_1>m_3)$~\cite{Tanabashi:2018oca}. A kinematic upper bound to the neutrino masses, mostly model independent, comes from the KATRIN experiment~\cite{KATRIN:2019yun,Aker:2021gma}, which measures the beta-decay spectrum of tritium atoms. KATRIN is sensitive to a linear combination of the neutrino masses; their most recent analysis yields an upper limit of $\sum_i\sqrt{|U_{ei}|^2 m_i^2}< 0.9\,{\rm eV}$ ($90\%$ confidence level)~\cite{Aker:2021gma}. The $U_{ei}$, $i=1,2,3$ elements of the mixing matrix are measured with good precision by neutrino oscillation experiments $|U_{e1}|^2\sim 0.7$, $|U_{e2}|^2\sim 0.3$, $|U_{e3}|^2\sim 0.02$ \cite{Esteban:2020cvm}.

The most stringent bounds on neutrino masses come from indirect measurements that rely on their effect on cosmological observables. Massless neutrinos are hot dark matter candidates and mediate a ``washing out'' of small-scale perturbations in the early Universe.
Observations of the Cosmic Microwave Background (CMB) by the Planck Satellite, combined with gravitational lensing data, Baryon Acoustic Oscillations (BAO) and large-scale structure limit the sum of neutrino masses $\sum_i m_i<0.13\,{\rm eV}$~\cite{DES:2021wwk}. Excluding BAO, this limit relaxes to $\sum m_i<0.24\,{\rm eV}$~\cite{Planck:2018vyg}. Adding Lyman alpha data and CMB temperature and polarization data to the lensing and the BAO data further improves the bound to $0.09\,{\rm eV}$~\cite{Palanque-Delabrouille:2019iyz}. 

On the theoretical front, extending the SM to incorporate neutrino masses has been a topic of intense research. The idea, if neutrinos are Majorana fermions, is to augment the SM in a way so as to generate the effective Weinberg operator $(LH)(LH)$~\cite{Weinberg:1979sa}, where $L$ is the lepton doublet containing the neutrino, and $H$ is the Higgs doublet. Popular mechanisms like the seesaw models, radiative mass models, and several others, rely on the introduction of new degrees of freedom at relatively high energy scales (see, for example, \cite{2016ARNPS..66..197D} and references therein). These new massive particles typically decouple from the cosmic plasma in the very early Universe, and hence do not alter its evolution. 

All current evidence of non-zero neutrino masses arises from experiments at redshift $z=0$. In some sense, the ``oldest" measurements of the non-zero nature of neutrino masses comes from solar neutrinos. Data from cosmology do not preclude a zero value of neutrino mass but only provide upper limits; the vanilla $\Lambda$CDM cosmology is perfectly consistent with zero neutrino masses~\cite{Planck:2018vyg}. As a result, scenarios where neutrinos are massless in the early Universe and gain mass only after recombination are not ruled out. Models predicting a late-time neutrino mass generation rely on time-varying neutrino masses arising out of the neutrino coupling to some time-varying scalar field~\cite{Fardon:2003eh,Krnjaic:2017zlz,Dev:2020kgz}, a late-time cosmic phase transition~\cite{Lorenz:2018fzb}, or the gravitational anomaly~\cite{Dvali:2016uhn}. 
Using a combination of CMB temperature and polarization power spectra, plus lensing data, the authors of~\cite{Lorenz:2018fzb} explored models where the neutrino masses are redshift dependent. They report a slight preference for models of late-time neutrino mass generated by a cosmic phase transition. In this scenario, due to the non-trivial dynamics of the phase transition, the bound on the current sum of neutrino masses is significantly weaker, $\sum m_\nu(z=0) < 4.8\,$eV at $95\%$CL. In a follow up work~\cite{Lorenz:2021alz}, the authors extracted the best-fit values of the neutrino masses as a function of redshift in a model-independent manner, using CMB and  BAO data and data from Type-IA SNe, and found a significantly weaker bound, $\sum_i m_{\nu_i}(z=0)<1.46\,{\rm eV}$ (95$\%$ CL). These looser bounds indicate, for example, that the hypothetical discovery of nonzero neutrino masses in future laboratory experiments \cite{Formaggio:2021nfz} would be consistent with bounds from cosmic surveys if we allow for late-time neutrino masses. 

The CMB probes high redshifts $(z\sim 1000)$, and one may wonder if there are other probes capable of testing the hypothesis that neutrino mass-generation occurs at much smaller redshifts. The answer to this question may lie in the diffuse supernova neutrino background (DSNB), a sea of MeV-neutrinos emerging from all supernova (SN) explosions in the Universe since the moment of the first stars $(z\lesssim 5)$~\cite{Lunardini:2005jf,Beacom:2010kk}. This isotropic, time-independent flux of neutrinos can be computed with precise knowledge of the underlying cosmology and the rate at which SNe happen in the Universe. The DSNB can be used as an excellent astrophysical laboratory to probe fundamental particle physics~\cite{DeGouvea:2020ang,Tabrizi:2020vmo,Das:2022xsz}.

The DSNB flux depends on whether the neutrinos are massive because of neutrino oscillations. For massless neutrinos, flavor eigenstates trivially coincide with mass eigenstates and will not undergo oscillations. However, the picture changes if the neutrinos acquire mass at a certain redshift. This leads to a scenario where the neutrino flavor and mass eigenstates are identical before a certain redshift (hence the mixing matrix is diagonal) and, as soon as they develop a non-zero mass, these two bases no longer coincide. This impacts the DSNB flux that arrives at the Earth in a nontrivial way. Neutrinos that were massless at the time of production would not suffer the usual effects that arise from neutrino oscillations inside the SN. As a result, we expect the net DSNB flux to be altered compared to what is predicted in the standard scenario.  

The detection of an altered DSNB flux can be used to probe such scenarios of late neutrino mass generation. The Super-Kamiokande (SK) experiment~\cite{Zhang:2013tua}, enriched with Gadolinium, is ready to search for the DSNB, and is expected to establish its existence within a decade~\cite{Super-Kamiokande:2021jaq}. Several upcoming experiments like Hyper-Kamiokande (HK)~\cite{Abe:2018uyc}, the Jiangmen Underground Neutrino Observatory (JUNO)~\cite{An:2015jdp}, and the Deep Underground Neutrino Experiment (DUNE)~\cite{Abi:2020evt} will also be instrumental in detecting the DSNB in the future. 
Moreover, the possibility of observing the total --all flavors-- DNSB flux via Coherent Elastic Neutrino-Nucleus Scattering (CEvNS) has been recently demonstrated in~\cite{Pattavina:2020cqc,Suliga:2021hek,Baum:2022wfc}.
As a result, the detection of the DSNB in the next few decades will serve as a unique probe of the epoch of neutrino mass-generation.

This work is organised as follows. We discuss our modelling of the DSNB flux in Sec.~\ref{sec:model}. We introduce our phenomenological approach to describing mass-varying neutrinos in Sec.~\ref{sec:massv}. In Sec.~\ref{sec:flux}, we determine the impact of late neutrino-mass generation on the DSNB fluxes to be measured on the Earth. We compute the expected event spectra in a DUNE-like detector in Sec.~\ref{sec:spectra}. We present our conclusions in Sec.~\ref{sec:conc}.
We use natural units where $\hbar = c = k_{\rm B} = 1$ throughout this manuscript.

\section{Modelling the DSNB flux}\label{sec:model}
\setcounter{equation}{0}
%
A prediction of the DSNB flux requires a good understanding of the evolution of the Universe, including the rate of core-collapse supernova (CCSN) $\R$, as well as a handle on the flavor-dependent neutrino spectra from a SN.
The CCSN rate, in turn, depends on the history of the star-formation rate (SFR), and has been measured by a number of independent astronomical surveys~\cite{Hopkins:2006bw}. The SFR data can be approximated by the following~\cite{Yuksel:2008cu,Horiuchi:2008jz}:
\begin{equation}\label{eq:SFR}
    \Dot{\r}_*(z)=\Dot{\r}_0 \left[(1+z)^{-10\,\al}+
   \left( \frac{1+z}{B}\right)^{-10\,\be}+ \left(\frac{1+z}{C}\right)^{-10\,\ga}\right]^{-1/10}\,,
\end{equation}
where $\Dot{\r}_0$ is the overall normalization of the rate and $\al,\be,\ga$ indicate the relevant slopes at different values of $z$. The parameters $B$ and $C$ are defined as 
\begin{eqnarray}\label{eq:SFRPar}
B&=&(1+z_1)^{1-\al/\be}\,,\\
C&=&(1+z_1)^{(\be-\al)/\ga}(1+z_2)^{1-\be/\ga}\,.
\end{eqnarray}
We quote the parameters for the SFR used in our work in Tab.~\ref{tab:sfr_params}. A more detailed discussion of these different parameters can be found in \cite{DeGouvea:2020ang}, and references therein. Using this, $\R(z)$ can be calculated as 
\begin{equation}\label{eq:CCSR}
    \R(z)=\Dot{\r}_*(z)\frac{\int_8^{100} \psi(M)\,dM}{\int_{0.1}^{100} M \psi(M)\,dM}\,,
\end{equation}
where $\psi(M)\propto M^{-2.35}$ -- the initial mass function (IMF) of stars -- gives the density of stars in a given mass range~\cite{Salpeter:1955it}. The lower limit on the IMF indicates tentatively the lowest mass at which a CCSN can form (we neglect lower mass electron capture SNe), while the upper limit is more ad hoc, including a reasonable fraction of failed SNe. 
\begin{table}[!t]
 \centering
 \caption{Star formation rate parameters and their uncertainties  used in this work.}\label{tab:sfr_params}
 \vspace{0.2cm}
 \begin{tabular}{|c|c|}
 \hline
 Parameter & Value\\
 \hline\hline
 $\dot{\rho}_0$ &  $0.0178^{+0.0035}_{-0.0036}~\mathrm{M_\odot~ y^{-1}~Mpc^{-3}}$\\
 $\alpha$ & $3.4\pm 0.2$\\
 $\beta$ & $-0.3 \pm 0.2$\\
 $\gamma$ & $-3.5\pm 1.0$\\
 $z_1$ & $1$\\
 $z_2$ & $4$\\
 \hline
 \end{tabular}
\end{table}

Finally, neutrino emission from a SN can be parameterized by the well-known alpha-fit spectra~\cite{Tamborra:2012ac}
\begin{align}\label{eq:Flux}
    F_{\nu_\beta}(E_\nu) = \frac{1}{E_{0\beta}}\frac{(1+\alpha)^{1+\alpha}}{\Gamma(1+\alpha)}\left(\frac{E_\nu}{E_{0\beta}}\right)^\alpha e^{-(1+\alpha)\frac{E_\nu}{E_{0\beta}}},
\end{align}
where $E_{0\beta}$ is the average energy for a flavor $\nu_\beta$ or $\overline{\nu}_{\beta}$, $\beta=e,\mu,\tau$ and $\alpha$ is a parameter that determines the width of the distribution. The DSNB spectra are dominated by neutrino emission 
from the cooling phase, where the spectra are approximately thermal. $\alpha=2.3$ approximates Eq.\,(\ref{eq:Flux}) as a Fermi-Dirac spectrum~\cite{Lunardini:2005jf}. 

\begin{figure}[!t]
\includegraphics[width=0.45\textwidth]{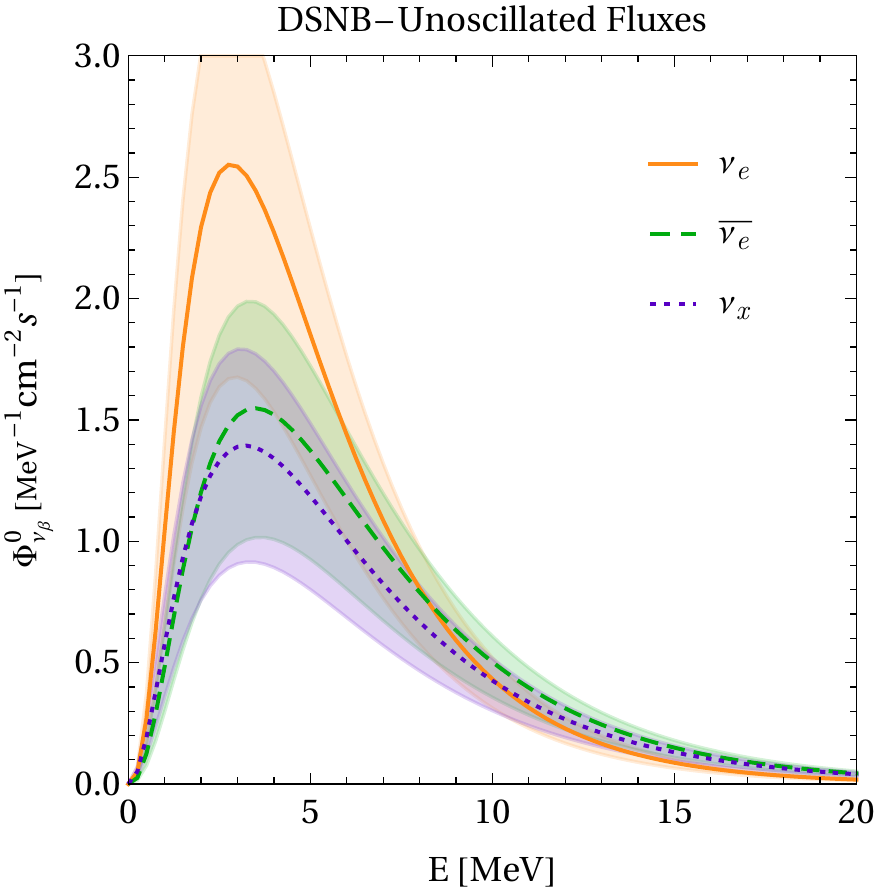}
\caption{Unoscillated DSNB flux $\Phi_{\nu_\beta}^0$ for each neutrino species, $\nu_e$ (orange), $\overline{\nu}_e$ (green dashed), and $\nu_x$ (purple dotted), as function of the neutrino energy $E$. The bands are associated to uncertainties in the star-formation rate.}
\label{fig:FluxNoOsc}
\end{figure}

With this information, in the absence of neutrino oscillations, the diffuse neutrino flux from all past SNe, is~\cite{Horiuchi:2008jz,Beacom:2010kk,Moller:2018kpn}
\begin{equation}\label{eq:DSNB}
\Phi_{\nu_\beta}^0(E)=\int_0^{z_{\rm max}}\frac{dz}{H(z)}\R(z) \phi_{\nu_\beta}^0(E(1+z))\,,
\end{equation}
where $H(z) = H_0\, \sqrt{\Omega_m (1+z)^3+\Omega_\Lambda}$ is the Hubble function with $H_0=67.36\,{\rm km\,s}^{-1}\,{\rm Mpc}^{-1}$ and $\Omega_m$, $\Omega_\Lambda$ represent the matter and vacuum contribution to the energy density, respectively~\cite{Aghanim:2018eyx}. The integral over $z$ ranges up to the maximum redshift of star-formation  $(z_{\rm max}\sim 5)$. 
$\phi_{\nu_\beta}^0(E)$ in Eq.~\eqref{eq:DSNB}, contains contributions from CCSNe and black-hole-forming (BHF) failed SNe,
\begin{align}
    \phi_{\nu_\beta}^0(E) = f_{\rm CC}\, F_{\nu_\beta}^{\rm CC}(E_\nu) + f_{\rm BH}\, F_{\nu_\beta}^{\rm BH}(E_\nu),
\end{align}
where $f_{\rm CC, BH}$ are the fraction of CC and BH-forming explosions, and $F_{\nu_\beta}^{\rm CC,BH}(E_\nu)$ are the time-integrated energy spectra for CCSNe and BHF-SNe. In the following, we take $f_{\rm BH}=21\%, f_{\rm CC}=1-f_{\rm BH}$. For the $F_{\nu_\beta}^{\rm CC, BH}(E_\nu)$, we have performed a fit of the time-integrated neutrino fluences obtained by the Garching group~\cite{Garching} in the form of Eq.~\eqref{eq:Flux}, taking as benchmark the data for $12 M_\odot$ for CCSNe and $40 M_\odot$ as for BHF-SNe.
We present in Fig.~\ref{fig:FluxNoOsc} the unoscillated fluxes at the Earth, obtained from Eq.~\eqref{eq:DSNB} for $\nu_e$ (orange), $\overline{\nu}_e$ (green dashed), and $\nu_x =\nu_\mu, \nu_\tau, \overline{\nu}_{\mu},\overline{\nu}_{\tau}$ and the corresponding antineutrinos (purple dotted). The $\nu_e$ flux is about twice the $\nu_x$ flux at $E\sim 3.5$~MeV. This difference arises mainly due to the interactions with neutrons that render the average energy of the $\nu_e$ flux smaller. Meanwhile, the $\overline{\nu}_e$ and $\nu_x$ have closer average energies, making the fluxes much more similar. Such difference will be crucial in our scenario of mass-varying neutrinos.

The neutrino flux gets processed through oscillation effects inside the SN and on the way to Earth. In this study, we neglect the effects of collective neutrino oscillations arising out of neutrino self-interactions deep inside the SN~\cite{Duan:2006an,Hannestad:2006nj}. The quantitative impact of collective oscillations is inconclusive to date, and we expect it to be relatively smaller for neutrinos predominantly produced in the cooling phase. The neutrino flux gets affected by adiabatic Mikheyev-Smirnov-Wolfenstein (MSW) resonant flavor conversion~\cite{PhysRevD.17.2369,Mikheev:1986gs}. Assuming NO, this implies that the $\nu_e$ are primarily emitted as $\nu_3$, while the non-electron neutrinos $\nu_{\mu,\tau}$ are emitted as combinations of $\nu_1$ and $\nu_2$. In this case, the final $\nu_e$ flux at the Earth $\Phi_{\nu_e}(E)$ is given by
\begin{equation}\label{eq:DSNBEarth}
\Phi_{\nu_e}(E)=|U_{e3}|^2\,\Phi_{\nu_e}^0 + (1-|U_{e3}|^2)\,\Phi_{\nu_x}^0,\,
\end{equation}
where $\Phi_{\nu_x}^0=\Phi_{\nu_\mu}^0=\Phi_{\nu_\tau}^0$, and $U_{\alpha i}$ is the Pontecorvo-Maki-Nakagawa-Sakata (PMNS) mixing matrix. 
Clearly, the DSNB flux depends on the underlying neutrino oscillation scenario. For example, if the neutrinos were massless at the time of the SN, the flavor evolution of the $\nu_e$ and $\nu_x$ (and the antineutrinos) would be trivial inside the explosion. On their way here, these would start oscillating, fast, once the neutrino masses turn on. In this case, the probability that a $\nu_{\alpha}$ is detected as a $\nu_e$ at the Earth is 
\begin{equation}
P_{\alpha e} = \sum_{i}|U_{\alpha i}|^2|U_{ei}|^2.
\label{eq:Pae_vac}
\end{equation}
Since SN neutrino energies are smaller than the muon mass, it is convenient to define $P_{xe}=P_{\mu e}+P_{\tau e}=1-P_{ee}$ so
\begin{equation}\label{eq:DSNBEarth_vac}
\Phi_{\nu_e}(E)=
P_{ee}\Phi_{\nu_e}^0 + 
(1-P_{ee})\,\Phi_{\nu_x}^0\,.
\end{equation}
In the next sections, we discuss in detail how the DSNB is modified if a fraction of it comes from neutrinos that were ``born'' with smaller masses or different mixing parameters. 

\section{Mass Varying neutrinos}\label{sec:massv}
\setcounter{equation}{0}
%
Following~\cite{Koksbang:2017rux,Dvali:2016uhn,Lorenz:2018fzb}, we assume that the neutrinos remain practically massless down to a certain redshift $z_s$ and gain a non-zero mass for $z<z_s$. We further assume the neutrino mass reaches its current value over a finite transition period. This could happen due to neutrinos coupling to the gravitational-$\theta$ term, causing a late phase transition in the Universe~\cite{Dvali:2016uhn}, or to neutrinos coupled to a scalar background, which evolves as a function of time~\cite{Fardon:2003eh,Berlin:2016woy,Krnjaic:2017zlz,Dev:2020kgz}.
Here, we remain agnostic regarding the details of mass-generation.

Assuming momentarily there is only one neutrino mass, we propose that it varies as a function of redshift according to
\begin{equation}\label{eq:NuMF}
    m_{\nu}(z)=\frac{m_{\nu}}{1+(z/z_s)^{{\rm B}_s}}\,,
\end{equation}
where $m_{\nu}$ is the current mass of the neutrino, ${\rm B}_s$ is a parameter which controls the width of the transition from a massless neutrino to a massive neutrino, and $z_s$ is the redshift below which the neutrino mass turns on. The specific form of the function is irrelevant and is chosen just to present a smooth transition to a non-zero mass. The values of $z_s$ and ${\rm B}_s$ determine when and at what rate the neutrino mass turns on.  

Since there are three neutrino masses, it is possible that they would ``turn on'' at different $z_s$ and that the transition would be associated to a different value of B$_s$. Here we assume a universal value for these two phenomenological paramaters. It is also possible to imagine that, as the neutrino mass turns on at $z_s$, so do all the PMNS mixing angles $\{\theta_{12},\theta_{13},\theta_{23}\}$, and that these turn on in a way that is also captured by Eq.\,(\ref{eq:NuMF}). We will discuss this possibility later but remain agnostic about the origin of such variations.

In the next section, we will detail the impact of redshift-dependent neutrino masses and mixing angles on the flavor evolution of neutrinos within the SN, as well as from the SN to Earth. 

\section{Impact of mass varying neutrinos on the DSNB}\label{sec:flux}
\setcounter{equation}{0}

\subsection{Only masses}

We first consider the case where the neutrino masses vary as a function of red-shift while the elements of the mixing matrix are time independent.

\subsubsection{Calculation of the survival probability}

In the standard three-massive-neutrinos paradigm, neutrinos produced via charged-current weak interactions are described as superpositions of the three neutrinos with well defined masses, $\nu_{\alpha} = U_{\alpha i}\nu_{i}$, $\alpha=e,\mu,\tau$, $i=1,2,3$. During propagation, the fact that the neutrino masses are different leads the neutrino flavor to oscillate; the associated oscillation lengths are inversely proportional to the differences of the squares of the neutrino masses. Global analysis of the present data indicate that the two independent mass-squared differences are $\Delta m^2_{21}\sim 7.5\times 10^{-5}\text{eV}^2$ and $\Delta m^2_{31}\sim \pm 2.5\times 10^{-3}\text{eV}^2$(plus for NO, minus for IO)~\cite{Esteban:2020cvm}. In matter, the neutrino flavor evolution is modified by the forward elastic neutrino--electron interaction amplitude along the neutrino path.\footnote{As mentioned earlier, we will ignore collective effects throughout.} This interaction is captured by a matter potential and modifies the effective Hamiltonian that describes neutrino flavor evolution~\cite{Wolfenstein:1977ue,Mikheyev:1985zog}. The matter potential depends on the electron number density ($n_{e}$) along the neutrino path. 
For position-dependent matter potentials, flavor evolution is rather involved. For certain matter profiles, however, the phenomenon is well understood~\cite{Parke:1986jy,Petcov:1987zj, Krastev:1988ci,Petcov:1988wv,Friedland:2000rn}.
The case where neutrinos are produced in a region of space where $n_e$ is large and propagate towards in the direction where $n_e$ falls roughly exponentially is well known and applies to both solar neutrinos and neutrinos produced in the core of SN explosions. 

In the limit where $G_Fn_e$, where $G_F$ is the Fermi constant, is much larger than $|\Delta m^2/E|$, where $E$ is the neutrino energy and $\Delta m^2$ are the neutrino mass-squared differences, electron neutrinos coincide with one of the propagation-Hamiltonian eigenstates (the one with the largest eigenvalue) in the production region. If neutrino flavor evolution is adiabatic inside the medium, the electron neutrino exits the matter distribution as a mass-eigenstate (eigenstate of the flavor-evolution Hamiltonian in vaccum). This ``mapping'' between the electron neutrino and mass-eigenstates depends on the mass ordering and whether we are considering electron neutrinos or antineutrinos, keeping in mind that the matter potential is positive for neutrinos, negative for antineutrinos.

Given what we know about the mass-squared differences, electron neutrinos, if the flavor evolution inside the SN is adiabatic, exit the SN as $\nu_3$ for NO and $\nu_2$ for IO. Electron antineutrinos, instead, exit the SN as $\overline{\nu}_1$ for NO and $\overline{\nu}_3$ for IO. In the adiabatic regime, it is easy to generalize this picture to the case where $G_Fn_e$ is not much larger than one or both $\Delta m^2/E$: the flavor evolution along the matter potential is just described by the effective mixing parameters at neutrino production.
\begin{figure}[!t]
\includegraphics[width=0.5\textwidth]{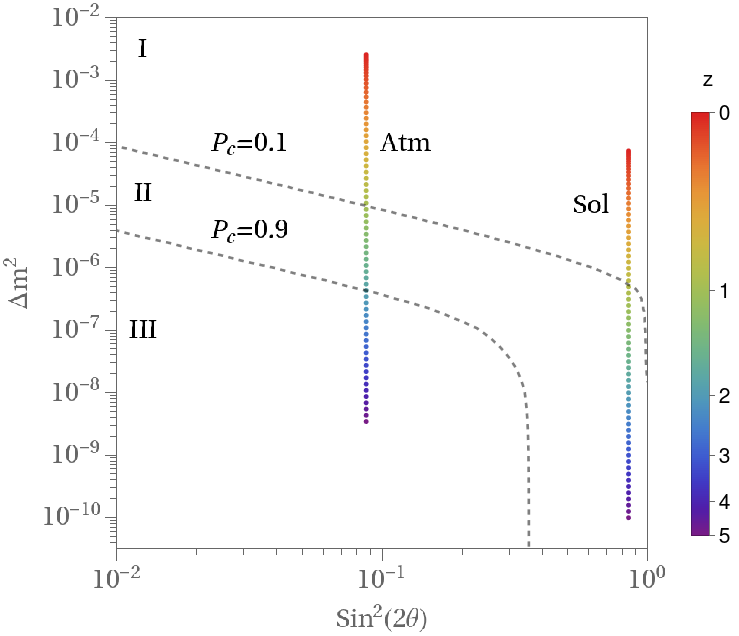}
\caption{Constant crossing probability contours in the $\sin^22\theta\times\Delta m^2$--plane. These define three regions: (I) $P_{c} < 0.1$, (II) $ 0.1 < P_c < 0.9$, and (III) $P_c > 0.9$. The color scale indicates the values of the two independent mass-squared differences as a function of the redshift of neutrino production. For the mass variation, we make use of Eq.~(\ref{eq:NuMF}) with $z_s = 0.32$ and $B_s = 5$.}
\label{fig:Adiab}
\end{figure}
In the case of two neutrino flavors, if the electron number density decreases roughly exponentially, adiabaticity is controlled by the ``crossing probability'' $P_{c}$. When $P_c$ vanishes, the flavor evolution is perfectly adiabatic. $P_c$ given by~\cite{Petcov:1987zj, Krastev:1988ci,Petcov:1988wv}
\begin{equation}
    P_{c} = \frac{\exp^{-(\pi\gamma F/2)} - \exp^{-(\pi\gamma F/2\sin^2\theta)}}{1-\exp^{-(\pi\gamma F/2\sin^2\theta)}},
\end{equation}
where F depends on the matter distribution inside the supernova and the mixing angle~\cite{Kuo:1988pn}. The dependence of $P_c$ on the oscillation parameters $\Delta m^2$ and $\theta$ is controlled by $\gamma$, which takes the following expression around the resonant region, defined by $n_e$ values that satisfy $2\sqrt{2}G_{F} E n_e = \Delta m^2\cos 2\theta$.
\begin{equation}
\gamma = \frac{\Delta m^2}{2E}\frac{\sin^2\theta}{\cos2\theta}\left(\frac{1}{n_e}\frac{dn_{e}}{dr}\right)^{-1}.
\end{equation}
If the variation of the effective mixing angles with the electron number density is slower than the oscillation wavelength in matter, $\gamma F >> 1$ and the neutrino evolution is adiabatic.  

It is easy to generalize the discussion to three flavors, taking advantage of the fact that the magnitudes of the two known mass-squared differences differ by two orders of magnitude. In this case, one can define two resonance regions and two crossing probabilities: $P_c^H$ ($H$ for high), associated to $\Delta m^2=\Delta m^2_{31}$ and $\theta=\theta_{13}$ and $P_c^L$ ($L$ for low), associated to $\Delta m^2=\Delta m^2_{21}$ and $\theta=\theta_{12}$. In our computations, in the standard case and NO, we use, for $P_c^H$, $\Delta m^2_{31} = 2.57 \times 10^{-3}\text{eV}^2$ and $\theta_{13} = 8.57^\circ$ and, for $P_c^L$, $\Delta m^2_{21} = 7.42 \times 10^{-5}\text{eV}^2$ and $\theta_{12} = 33.44^\circ$~\cite{Esteban:2020cvm}.

The flavor-at-production and the neutrino spectrum emitted by the supernova depends on the evolution of the collapse of the star. After the shock-wave, free electrons are captured by free protons generated by the dissociation of nuclei yielding a $\nu_{e}$-rich flux -- the neutronization burst. Thereafter, a large fraction of the neutrinos are emitted during the cooling phase when the supernova loses the remaining gravitation binding energy via the thermal emission of neutrinos of all flavors. In this phase, the temperature of $\nu_e$ is expected to be smaller than that of $\overline{\nu}_e$ and that of $\nu_x$, since $\nu_e$ interacts more strongly with the production medium. 
%
%
Most of the neutrinos are created deep inside the explosion where the density is quite large. On their way out, neutrinos cross both the atmospheric ($\rho\sim 3\times 10^{3}~{\rm g/cm^3}$), and the solar resonances ($\rho\sim 40~{\rm g/cm^3}$) at lower densities. Both resonances happen well outside of the neutrinospheres. In Fig.~\ref{fig:Adiab}, we depict contours of constant $P_c$ in the $\Delta m^2\times\sin^2 2\theta$-plane. We identify three qualitatively distinct regions: (I) $P_c < 0.1$, where flavor-evolution ``through'' the resonance is adiabatic, (II) $0.1 < Pc < 0.9$ (region II), and (III) $P_c > 0.9$, where neutrino flavor-evolution is highly non-adiabatic. Given the current values of the mass-squared differences ($z=0$ in the figure), flavor-evolution is very adiabatic through both the atmospheric and solar resonances~\cite{Dighe:1999bi}. The large value of the density in the region where the neutrinos are produced 
leads to, as discussed earlier, $\nu_e$ being mapped to the most massive state (e.g.~$\nu_3$ for NO) while the $\nu_x$ is mapped into the lighter states (e.g., for NO, some combination of $\nu_1$ and $\nu_2$). In this case, for NO, the flux of electron neutrinos at the Earth is given by the projection of the three massive states weighted by the initial flux -- Eq.~(\ref{eq:DSNBEarth}). The small value of $|U_{e3}|^2\sim 0.02$ implies that most of the $\nu_{e}$ at the Earth started out as a $\nu_x$ deep inside the explosion.

If the neutrino masses were smaller at a given time in the history of the Universe, the flavor evolution inside the supernova might no longer be adiabatic. This is depicted in Fig.~\ref{fig:Adiab}, where we indicate the different values of the mass-squared differences for different redshifts, following Eq.~(\ref{eq:NuMF}) with $z_s = 0.32$ and $B_s = 5$. This allows the possibility that one massive state ``flip'' into another one as the neutrinos propagate through the supernova. In the case of a non-adiabatic evolution and NO, the initial $\nu_e$ component of the flux will also be partially mapped to $\nu_1$ and $\nu_2$ fluxes outside the SN. Following~\cite{Dighe:1999bi}, the $\nu_e$ flux at the Earth in the case of a non-adiabatic evolution is given by Eq.~(\ref{eq:DSNBEarth_vac}) where
%
%
the $\nu_e$ survival probability $P_{ee}$ is given by
\begin{equation}
 P_{ee}=|U_{e1}|^2P^H_{c}P^L_{c} + |U_{e2}|^2(P^H_c - P^H_cP^L_c) + |U_{e3}|^2(1 - P^H_c).
\end{equation}
In the adiabatic limit ($P^L_c = P^H_c = 0$), we recover the standard expression for the $\nu_e$ flux at the Earth, Eq.~(\ref{eq:DSNBEarth}). As the neutrino masses decrease, the atmospheric and solar resonances shift to lower densities. Note that, if the neutrino mass is low enough, the neutrinos might not ``cross'' one of the resonances on their way out of the SN. That will also impact the final $\nu_e$ flux. 

\begin{figure}[!t]
\includegraphics[width=0.5\textwidth]{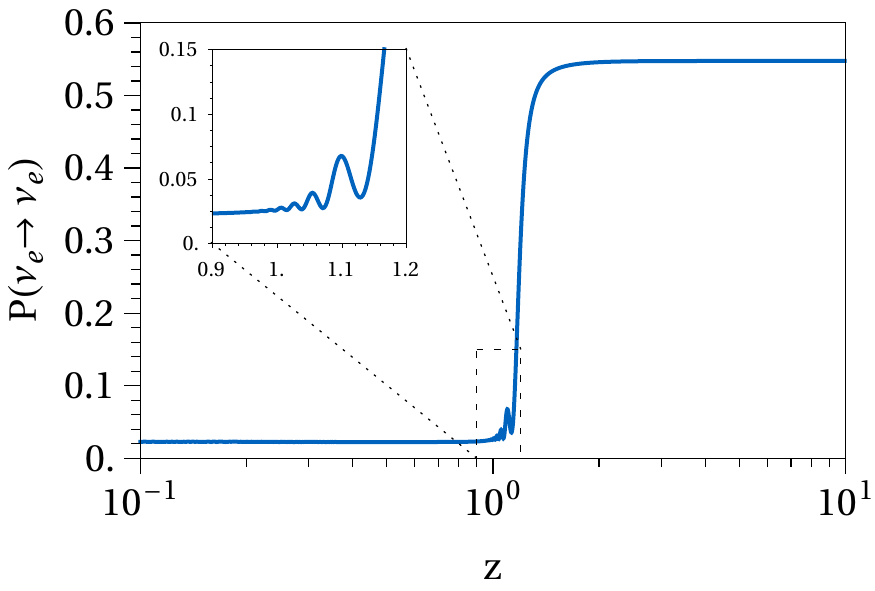}
\caption{Electron-neutrino survival probability as function of the redshift for $E_{\nu} = 10$~MeV. We consider that the mass changes as a function of the redshift according to Eq.~(\ref{eq:NuMF}) for $z_s = 0.32$ and $B_s = 5$.}
\label{fig:Pee}
\end{figure}

In the case of the normal mass ordering, the non-adiabatic evolution leads to an enhancement of the $\nu_e$ flux because the initial $\nu_e$ flux is larger than that of the other flavors. Fig.~\ref{fig:Pee} depicts the electron-neutrino survival probability on the Earth as a function of the redshift $z$ of the SN, for $E_{\nu} = 10$~MeV, $z_{s} = 0.32$ and $B_{s} = 5$. For this choice of mass-varying parameters, the transition between massless and massive neutrinos happens around $z \sim 1$. If the neutrino energy increases, the transition shifts to lower redshifts. Around $z\sim 1$, we observe a small oscillatory pattern in $P_{ee}$, highlighted in the inset. For those values of $z$, $\Delta m^2_{21} \sim 10^{-8}$~eV$^2$ and the associated oscillation length is of order the size of the SN.  

\subsubsection{The DSNB $\nu_e$ flux on Earth}

In order to include the possibility that neutrino masses are redshift-dependent, Eq.~\eqref{eq:DSNBEarth} needs to be altered: 
\begin{subequations}
\begin{align}\label{eq:DSNB_mvz}
\Phi_{\nu_e}(E)&=\int_0^{z_{\rm max}}\frac{dz}{H(z)}\R(z) \left\{P_{ee}(z) \phi_{\nu_e}^0 + (1-P_{ee}(z))\phi_{\nu_x}^0\right\}\,,\\
\Phi_{\bar{\nu}_e}(E)&=\int_0^{z_{\rm max}}\frac{dz}{H(z)}\R(z) \left\{\overline{P_{ee}}(z) \phi_{\bar{\nu}_e}^0+ (1-\overline{P_{ee}}(z))\phi_{\nu_x}^0\right\}\,,\\
\Phi_{\nu_x}(E)&=\int_0^{z_{\rm max}}\frac{dz}{H(z)}\R(z) \frac{1}{4}\left\{(1-P_{ee}(z)) \phi_{\nu_e}^0 +(1-\overline{P_{ee}}(z)) \phi_{\bar{\nu}_e}^0 + (2+P_{ee}(z)+\overline{P_{ee}}(z))\phi_{\nu_x}^0\right\}\,,
\end{align}
\end{subequations}
where, for clarity, we omitted the dependence of $\phi^0_{\nu_\beta}$ on $(E,z)$. $P_{ee}(z)$ ($\overline{P_{ee}}(z)$) indicate the oscillation probabilities for neutrinos (antineutrinos) from a SN explosion at a redshift $z$ as described in the last subsection. 
It depends on the ${\rm B}_s,z_s$ parameters, so the final DSNB flux will contain information regarding them. 
The DSNB is an integrated flux, so, in principle, there is no an explicit way to distinguish neutrinos that were emitted at higher redshifts from those produced more recently. 
However, the energies of the neutrinos produced earlier are more redshifted and hence we expect that time-dependent neutrino masses will distort the DSNB energy spectrum.
\begin{figure}[!t]
\includegraphics[width=\textwidth]{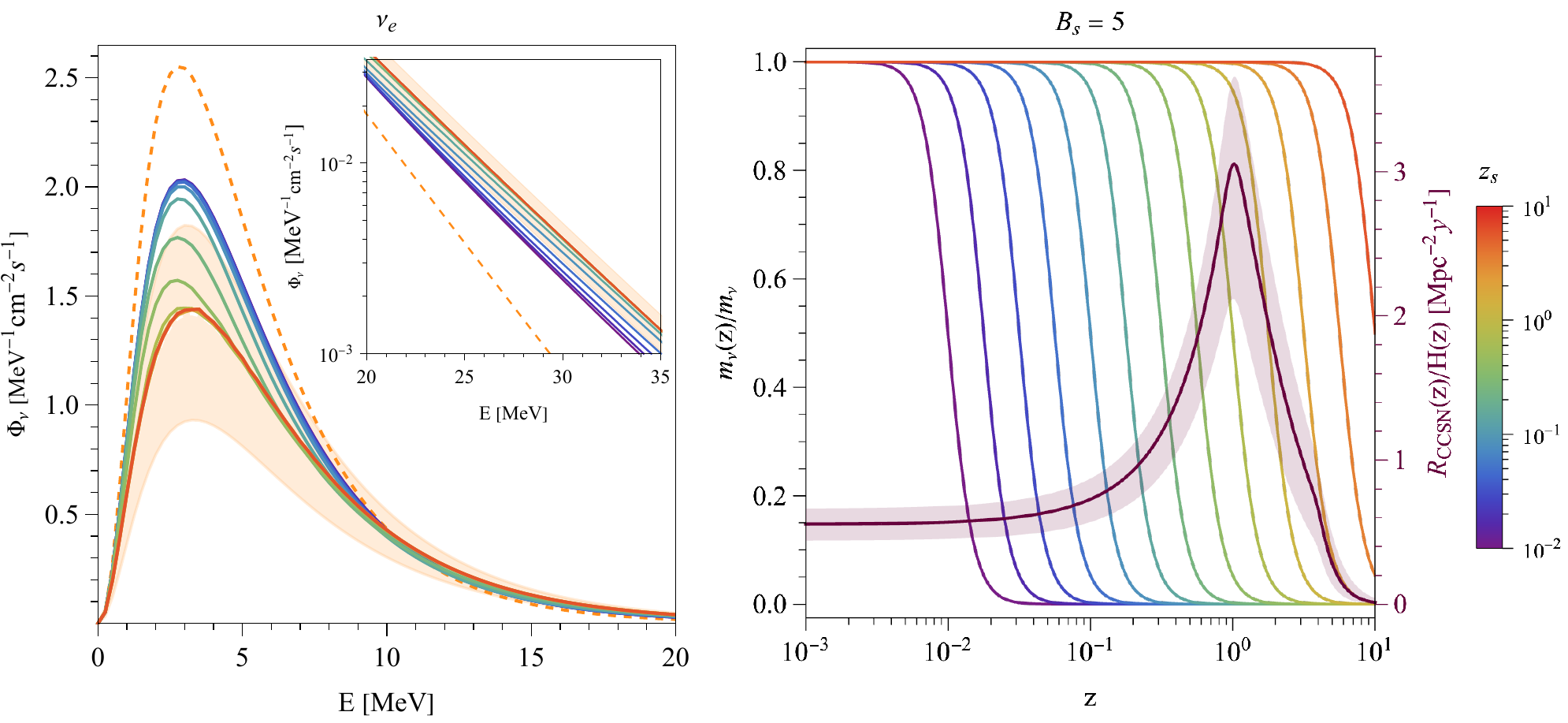}
\caption{(Left) The DSNB $\nu_e$ flux as a function of the neutrino energy for different values of $z_s\in [10^{-2},1]$ (rainbow colored) and for the standard case (orange) including the star-formation-rate uncertainty (orange band). The dashed orange line indicates the DSNB flux assuming massless neutrinos, i.e., $P_{ee}(z) = 1$ for all $z$. (Right) Neutrino mass as a function of redshift $z$, normalized to the current value of the mass, $m_{\nu}(z)/m_\nu$, together with $\R(z)/H(z)$. See text for details. In both panels, ${\rm B}_s=5$. We assume the normal ordering for the neutrino masses.}
\label{fig:F_mvz_1}
\end{figure}

We first consider the case where only the neutrino masses change over time, assuming the NO. Fig.~\ref{fig:F_mvz_1}, left, depicts the electron neutrino flux at the Earth for different values of $z_s\in [10^{-2},1]$, for ${\rm B}_s=5$. 
The standard flux, for constant neutrino masses, including uncertainties associated to the SFR, is depicted as the orange band while the dashed orange line corresponds to the unoscillated DSNB flux, $\Phi^0_{\nu_e}$ (i.e., expectations in the scenario where all neutrino masses are exactly zero). 
We observe that the hypothesis that neutrino masses depend on the redshift can significantly impact the DSNB electron-neutrino flux for $z_s\lesssim 0.5$. 
In fact, for $E=3~\MeV$, the DSNB flux can be larger than standard expectations by a factor of order $1.4$ for $z_s\sim 10^{-2}$. Moreover, from a simple flux conservation argument, this also implies that the flux at larger energies is reduced with respect to the standard case, see the inset plot. Such neutrinos would have acquired their masses rather recently, when the Universe was $13.652~{\rm Gyr}$ old (compare with the age of the Universe $t_0=13.795~{\rm Gyr}$), therefore the DSNB was mostly produced when neutrinos were virtually massless.
The increment on the $\nu_e$ flux at low energies is directly related to the difference between the unoscillated $\nu_e$ and $\nu_x$ fluxes. In the standard scenario $P_{ee}\ll 1$, so the $\nu_e$ flux at the Earth is basically the $\nu_x$ flux produced at the neutrinosphere, which is much broader in energy. However, if $P_{ee}$ significantly differs from the standard case, the contribution from the $\nu_e$ flux that exited the neutrinosphere becomes significant, thus modifying the $\nu_e$ flux at the Earth.

For values of $0.1\lesssim z_s\lesssim 1$, the $\nu_e$ flux is still larger than the SM flux at low energies. 
Meanwhile, for $z_s\gtrsim 1$, the DSNB flux is basically indistinguishable from the standard case.
To understand the dependence on the values of $z_s$, we show in the right panel of Fig.~\ref{fig:F_mvz_1} the redshift evolution of neutrino masses along with the factor $\R(z)/H(z)$, cf.~Eq.~\eqref{eq:DSNB}.
This object describes the SN neutrino production as a function of redshift, including effects associated to the expansion of the Universe. It reveals that most of the DSNB flux is produced at $0.1\lesssim z\lesssim 5$.
%
Thus, if $z_s\gtrsim 1$, the SNe matter effects are basically the same as in the standard case, so we do not expect any impact from the mass-varying hypothesis.
On the other hand, if $0.1\lesssim z_s\lesssim 1$, a significant fraction of the  DNSB comes from SN explosions that happened when the neutrino masses were significantly smaller. The largest effects occur when neutrinos were effectively massless during most of the history of the Universe, $z_s\lesssim 0.1$, as noted above.

\begin{figure}[!t]
\includegraphics[width=\textwidth]{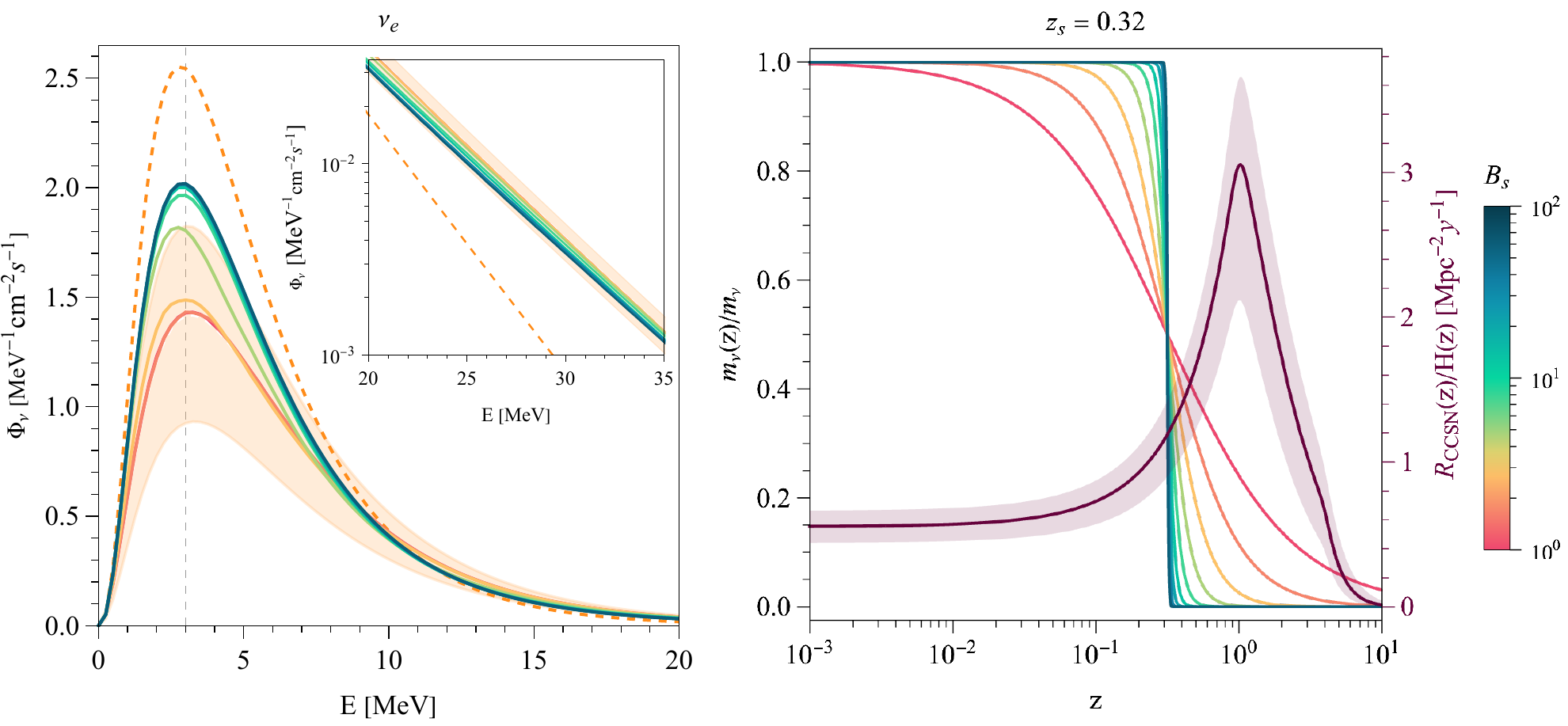}
\caption{Same as Figure~\ref{fig:F_mvz_1}, for different values of ${\rm B}_s\in[1,100]$ and fixed $z_s=0.32$.}
\label{fig:F_mvz_2}
\end{figure}

Fig.~\ref{fig:F_mvz_2} captures the dependence of the DSNB flux on the parameter ${\rm B}_s$ for a fixed $z_s=0.32$. 
${\rm B}_s$ controls how fast neutrino masses increase, larger values associated to more abrupt transitions. For $B_s\gtrsim 10$, the transition is almost instantaneous.
Whenever the growth of neutrino masses is rapid (${\rm B}_s\gtrsim 50$) the flux is relatively larger (by the same factor, discussed earlier, $\Phi_{\nu_e}/\Phi_{\nu_e}|_{\rm SM}\sim 1.4$).
This dependence on ${\rm B}_s$ again is understood by comparing the redshift dependence of both $m_\nu(z)$ and $\R(z)/H(z)$ (right panel).
If the masses become non-zero instantaneously, neutrinos emitted before the transition ($z>z_s$) would have been effectively massless, and their contribution to the $\nu_e$ flux  will be associated to the electron neutrino survival probability $P_{ee}=\sum_k|U_{ek}|^4\approx 0.57$, characteristic of electron neutrinos propagating very long distances in vacuum. 
After the masses ``turn on,'' neutrinos will be subject to matter effects inside the SN and $P_{ee}=|U_{e3}|^2$ in the NO, as discussed earlier.
The final DSNB flux will be an amalgam of neutrinos from two different epochs whose contributions are weighted by the SFR divided by the expansion rate.
If the transition is not instantaneous (small B$_s$), the DSNB flux is reduced because matter effects would impact the propagation inside the SN for a longer period of time and the neutrino masses would be of order the current masses for an extended range of redshifts (when the neutrino masses are $\gtrsim 10\%$ of the masses today, the matter effects are very similar to the standard case.). 
Thence, the DSNB flux in such cases is closer to the standard case, as can be observed in Fig.~\ref{fig:F_mvz_2}.

\begin{figure}[!t]
\includegraphics[width=\textwidth]{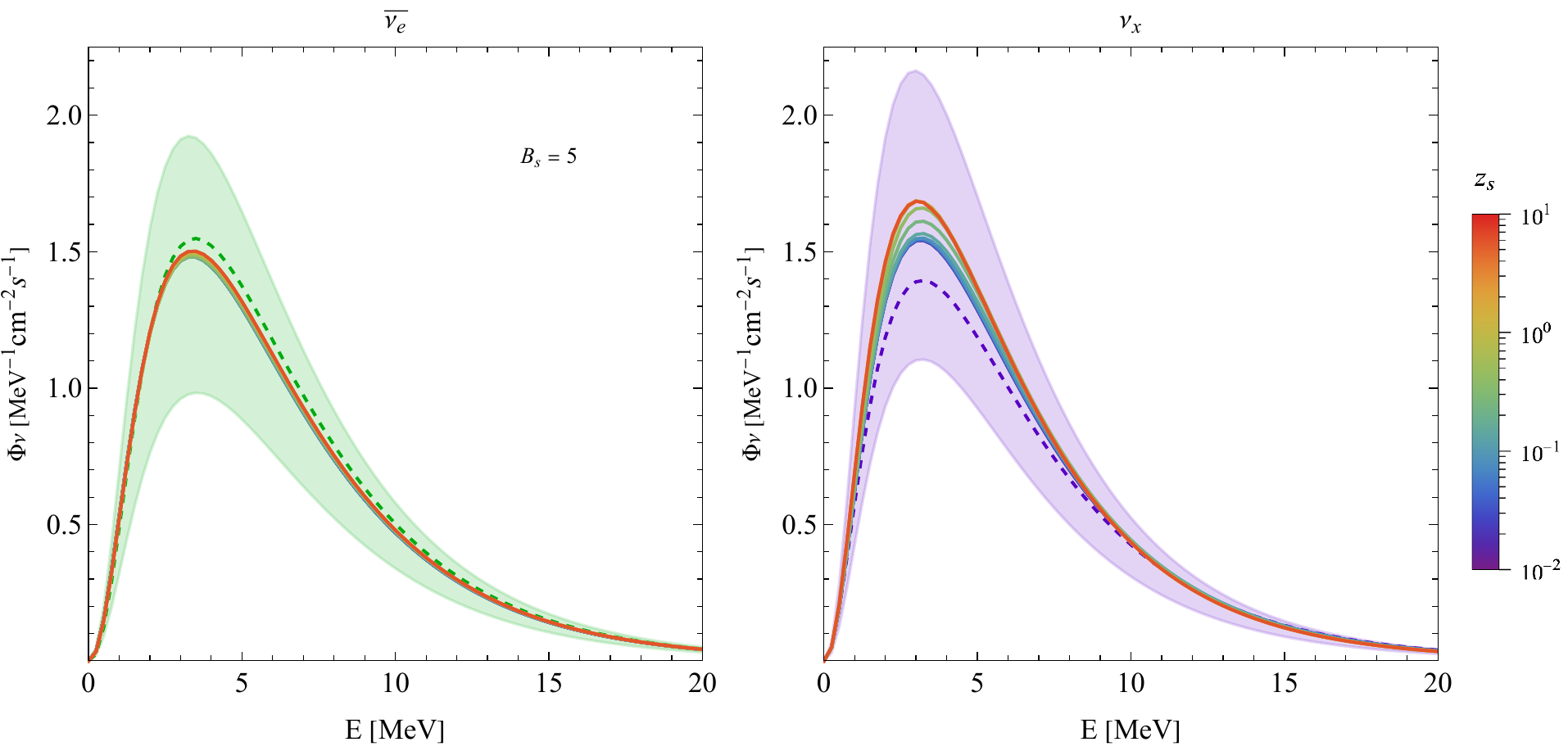}
\caption{The DSNB $\overline{\nu}_e$ flux (left) and $\nu_x$ flux (right) as a function of the neutrino energy for different values of $z_s\in [10^{-2},1]$ (rainbow colored) and fixed ${\rm B}_s=5$. The fluxes in the standard case, including the star-formation-rate uncertainty, define the green (left) and purple (right) bands. The dashed lines indicate the DSNB flux assuming massless neutrinos. We assume the normal ordering for the neutrino masses. 
}
\label{fig:F_mvz_3}
\end{figure}

So far, we have focused on the impact of redshift-dependent neutrino masses on the $\nu_e$ flux assuming NO. Instead the impact on the $\overline{\nu}_e$ spectrum is minimal. This is depicted in Fig.~\ref{fig:F_mvz_3}, left, for NO. This indifference is not strongly dependent on the mass ordering and is mostly a consequence of the fact that the original $\overline{\nu}_e$ and $\nu_x$ fluxes (keeping in mind that $\nu_x$ includes the antineutrino flavors) are very similar, see Fig.~\ref{fig:FluxNoOsc}. In this case, oscillation effects are invisible. Nonetheless, it is worth discussing the oscillation of $\overline{\nu}_e$ in a little more detail.   
The standard prediction assuming adiabatic propagation indicates that $\overline{\nu}_e$ emerges from the SN as $\overline{\nu}_1$ so $\overline{P_{ee}}=|U_{e1}|^2\approx 0.67$ at low energies. The $\overline{\nu}_e$ flux at the Earth is, therefore, roughly an equal admixture of $\Phi_{\overline{\nu}_e}^0$ and $\Phi_{\nu_x}^0$, fluxes that are close to each other.
Furthermore, if neutrino masses arise later in the evolution of the Universe ($z_s\lesssim 0.1$),  $\overline{P_{ee}}\sim 0.57$ as in the $\nu_e$ case. 
The difference between these fluxes is safely within the star-formation-rate uncertainty, making the effect unobservable, even in the most optimistic cases. Similarly, 
$\nu_x$, measurable only via NC interactions at these low energies, is also modified in a negligible way, as can be observed in the right panel of Fig.~\ref{fig:F_mvz_3}, for NO. Since it contains the contributions of both neutrinos and antineutrinos, the modification of this flux is at most $10\%$, within the star-formation-rate uncertainty.

For the IO, the situation changes significantly for $\nu_e$. In the standard scenario, the MSW effect predicts that the a $\nu_e$ created at the neutrinosphere leaves the SNe as a $\nu_2$ mass eigenstate, so $P_{ee}=|U_{e2}|^2\sim 0.3$. 
Meanwhile, if neutrinos only acquired their masses recently ($z_s\lesssim 1$), we would have the same probability as in the NO, $P_{ee}=\sum_k|U_{ek}|^4\approx 0.57$. 
Thus, we find that any possible modification on the DSNB energy spectrum in the IO will lie within the current star-formation uncertainty band.  
On the other hand, for antineutrinos and the IO, we have $\overline{P_{ee}}=|U_{e3}|^2$, so, if matter effects inside the SNe were significantly different at some point of the evolution of the Universe, $\overline{P_{ee}}$ would be considerably different from the standard value. 
Nevertheless, since $\Phi^0_{\nu_x}$ and $\Phi^0_{\overline{\nu}_e}$ are very similar, cf. Fig.~\ref{fig:FluxNoOsc}, any imprint of the mass-varying hypothesis would be very difficult to measure.

\subsection{Masses and mixing}

If one allows for the possibility that the neutrino masses are redshfit-dependent, it is reasonable to ask whether the the neutrino mixing parameters also depend on the redshift. We address this possibility in this subsection. 

\subsubsection{Calculation of the survival probability}

\begin{figure}[!t]
\includegraphics[width=0.5\textwidth]{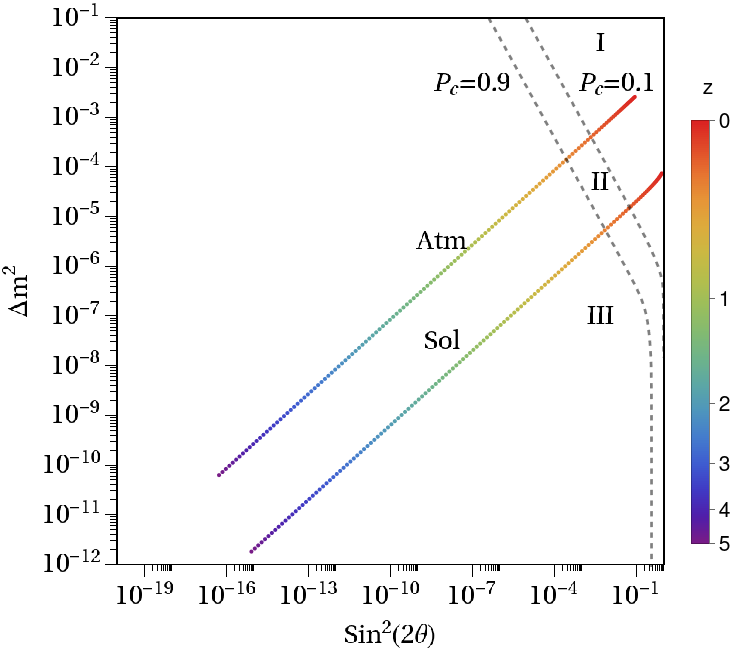}
\caption{Constant crossing probability contours in the $\sin^22\theta\times\Delta m^2$--plane. These define three regions: (I) $P_{c} < 0.1$, (II) $ 0.1 < P_c < 0.9$, and (III) $P_c > 0.9$. The color scale indicates the values of the two independent sets of oscillation parameters -- $\Delta m^2_{31}$ and $\sin^22\theta_{13}$ (Atm) and $\Delta m^2_{21}$ and $\sin^22\theta_{12}$ (Sol) -- as a function of the redshift of neutrino production. For the mass variation, we make use of Eq.~(\ref{eq:NuMF}) with $z_s = 0.32$ and $B_s = 5$.}
\label{fig:AdiabMM}
\end{figure}

In the case where both the neutrino masses and the mixing parameters depend on the redshift, the adiabaticity of the neutrino flavor-evolution inside the SN is modified relative to the case where only the masses depend on the redshift. Similar to Fig.~\ref{fig:Adiab}, Fig.~\ref{fig:AdiabMM} depicts contours of constant $P_c$ in the $\Delta m^2\times \sin^22\theta$-plane along with the $z$-dependent values of the oscillation parameters. Here, however, both the masses and mixing angles go to zero as $z$ grows. Explicitly, we postulate that the redshit-dependent mixing angles $\theta_{ij}$, $ij=12,13,23$  are
\begin{equation}\label{eq:thMF}
    \theta_{ij}(z)=\frac{\theta_{ij}}{1+(z/z_s)^{{\rm B}_s}}\,.
\end{equation}
Similar to the mass varying scenario, the non-adiabatic evolution of the neutrinos for smaller masses and mixing angles will lead to an enhancement of the $\nu_e$ flux in the Earth. 
The lower values of the mixing parameters imply that the flavor resonances happen at lower electron number densities. The minimum densities considered here are around $ 2\g/{\rm cm^3}$. If the MSW resonance happens at lower densities, neutrinos will not ``cross'' them as they exit the supernova.
In this case, flavor-evolution resembles the vacuum case~\cite{Friedland:2000cp}. For $B_s = 5$ and $z_{s} = 0.32$, this happens for $z\sim 1.3$ for the ``atmospheric'' resonance and $z\sim 0.6$ for the solar one.

\subsubsection{The DSNB $\nu_e$ flux at Earth}
 
\begin{figure}[!t]
\includegraphics[width=\textwidth]{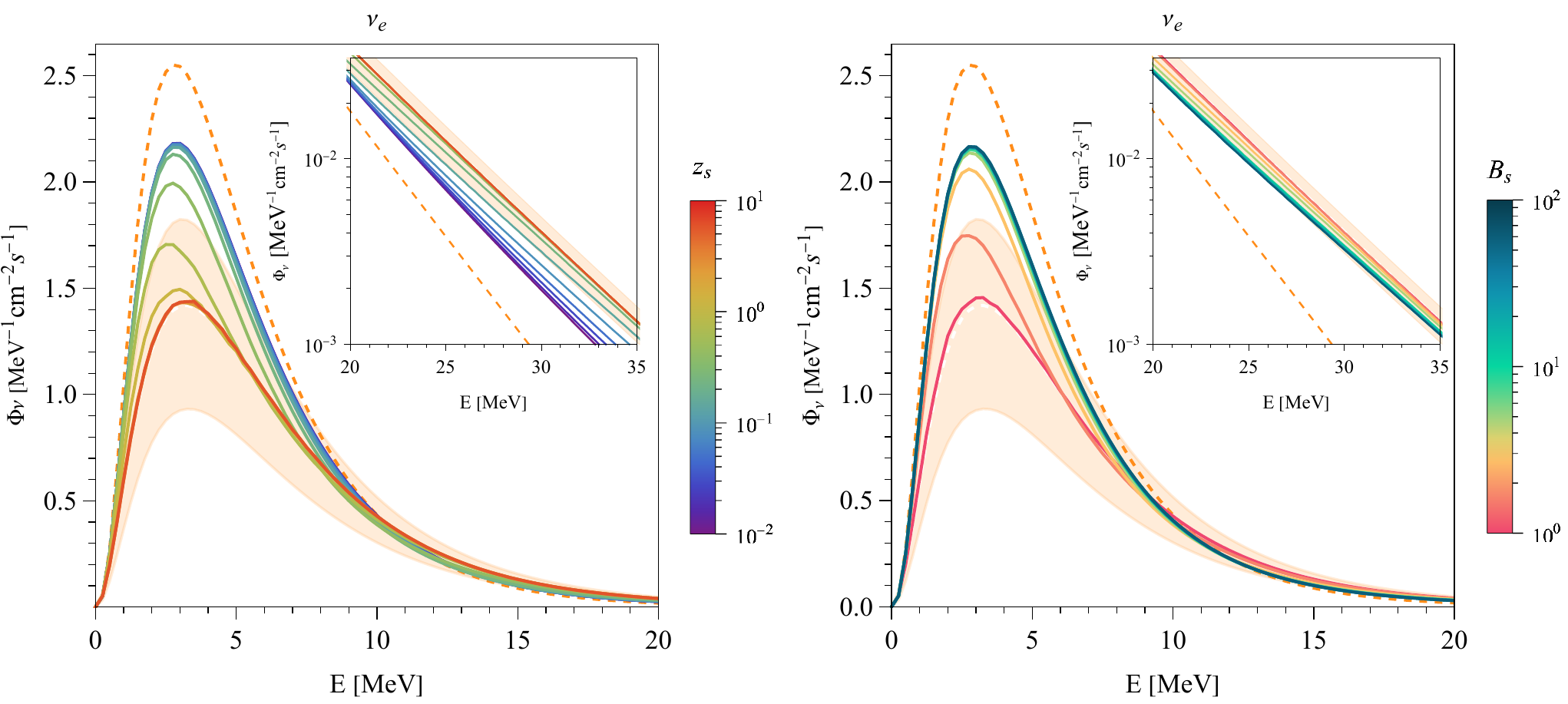}
\caption{(Left) The DSNB $\nu_e$ flux as a function of the neutrino energy for different values of $z_s\in [10^{-2},1]$ (rainbow colored) and fixed ${\rm B}_s=5$ and for the standard case (orange) including the star-formation-rate uncertainty (orange band). The dashed orange line indicates the DSNB flux assuming massless neutrinos, i.e., $P_{ee}(z) = 1$ for all $z$. (Right) The DSNB $\nu_e$ flux as a function of the neutrino energy for different values of ${\rm B}_s\in [1,100]$ (rainbow colored) and fixed $z_s=0.32$ and for the standard case (orange) including the star-formation-rate uncertainty (orange band). The dashed orange line indicates the DSNB flux assuming massless neutrinos, i.e., $P_{ee}(z) = 1$ for all $z$. We assume the normal ordering for the neutrino masses.}
\label{fig:F_mvz_VOP}
\end{figure}

We compute the DNSB flux as discussed around Eq.~\eqref{eq:DSNB_mvz}, this time including in the $z$-dependency of $\theta_{ij}(z)$ ($ij=\{12,13,23\}$).
As in the previous subsection, we concentrate on the NO and on electron neutrinos, where we anticipate the strongest effects.
Similar to the results presented in the last subsection, Fig.~\ref{fig:F_mvz_VOP} depicts the DSNB $\nu_e$ flux as function of the neutrino energy for different values of $z_s, {\rm B}_s$. 
In the left panel, we fix ${\rm B}_s=5$, and vary $z_s\in [10^{-2},10]$. In the right panel, we fix $z_s=0.3$, and vary ${\rm B}_s\in [1,100]$. We observe that the enhancement of $\Phi_{\nu_e}/\Phi_{\nu_e}|_{\rm SM}\sim 1.5$ at $E=3\MeV$, larger than what we found in the mass-only varying case.
At higher energies, instead, the flux is relatively suppressed, by a factor ranging from, roughly, $0.6$ for $E=20\MeV$ to $0.4$ at $E=50\MeV$. Hence, the fact that the mixing angles also decrease with increasing redshift leads to more pronounced effects.
Taking as example $z_s=0.05, {\rm B}_s=5$, we find that a $\nu_e$ emitted at redshifts $z\gtrsim z_s$ will exit mostly as a $\nu_1$ (assuming the normal mass ordering) since the mixing angles are small enough that the PMNS matrix is effectively diagonal.
After exiting the supernova, neutrino masses turn on in such a way that they remain a $\nu_1$ throughout.
Thus, at the Earth, the electron survival probability is simply $P_{ee}=|U_{e1}|^2\approx 0.67$, which enhances the observable $\nu_e$ flux at lower energies. 
For the same reason, the flux is suppressed at higher energies $E\gtrsim 10$~MeV, such that, for $E\gtrsim 20\MeV$, the flux lies below the smallest value allowed by the uncertainty on the SFR, see the inset in the left panel.

The dependence on $z_s$ and ${\rm B}_s$ of the final flux is similar to the masses-only varying case.
If $z_s\lesssim 0.1$, the DSNB is mostly composed of neutrinos that were emitted when their masses and mixing angles where small.
On the other hand, if $z_s\gtrsim 2$, the largest contribution to the DSNB comes from neutrinos produced with masses and mixing angles similar to the ones observed today.
In the latter case, the DSNB will be consistent with standard values.
On the other hand, depending on how fast the transition between almost massless neutrinos and the observed mixing pattern occurs, parametrized by ${\rm B}_s$, the flux is enhanced at low energies.
If the transition is rather sharp ($B_s\gtrsim 10$), the DSNB is simply the superposition of a nearly massless component coming from SN explosions with $z>z_s$, and a standard part emitted when $z<z_s$.
For smaller values of ${\rm B}_s$, the dependence on redshift is smoother, leading to a small variation of the masses as a function of  redshift.
For instance, for ${\rm B}_s = 1$, $\Delta m_{ij}^2(z=10)/\Delta m_{ij}^2\sim 0.02$ and the propagation in the SN is still adiabatic. In this case, there are no significant changes to the DSNB spectrum.

The $\overline{\nu}_e$ spectra in this case is virtually unaltered. 
The MSW adiabatic flavor conversion predicts that $\overline{P_{ee}}=|U_{e1}|^2$, value equal to the probability obtained when the mixing angles are small.
Since $\overline{\nu}_e$ would be mostly composed by $\overline{\nu}_1$, because the PMNS matrix would be close to diagonal, the predicted antineutrino flux at the Earth would be identical to the standard case.
As before, a measurement of both neutrinos and antineutrinos from the DNSB would be crucial to test this scenario.

%
\section{Event Spectra in a DUNE-like detector}\label{sec:spectra}
\setcounter{equation}{0}
%

The detection of the DSNB is one of the main goals of current and future experiments, including SK and HK~\cite{Super-Kamiokande:2021the,Hyper-Kamiokande:2018ofw}, JUNO~\cite{JUNO:2015zny}, and DUNE~\cite{DUNE:2020lwj}. 
Our results from the last section can be summarized as follows. If the neutrino mass varies as a function of redshift around $z\sim 1$ and if the neutrino mass ordering is normal, we expect the DNSB $\nu_e$ flux to be very different from standard expectations. The DSNB $\overline{\nu}_e$ flux, on the other hand, is quite indifferent to the potential $z$-dependency of neutrino masses.  
These two facts point towards a simple strategy for testing the hypothesis that neutrino masses ``turn on'' as a function of time. The detection of the DSNB $\overline{\nu}_e$ flux in experiments like SK and HK,\footnote{These experiments, along with scintillator experiments, predominantly detect the DSNB via inverse beta-decay.} can be used to normalize the total flux, thus reducing systematic uncertainties, including those related to uncertainties in the SFR.
Meanwhile, data from an experiment like DUNE, which can detect electron neutrinos instead of antineutrinos, can be used to provide information on whether the $\nu_e$ spectrum is consistent with standard expectations.

\begin{figure}[!t]
\includegraphics[width=0.45\textwidth]{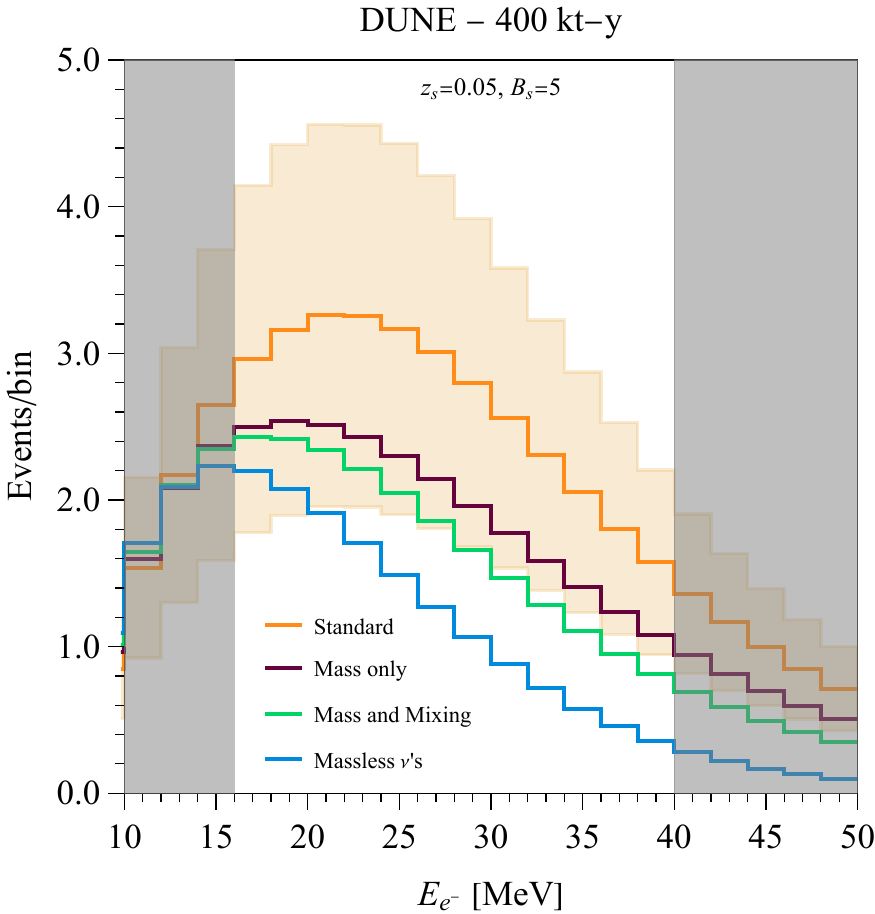}
\caption{$\nu_e$ DSNB event spectra in a DUNE-like detector, assuming 400 kton-years of exposure, as function of the recoil-electron energy. The spectrum in the standard case is in orange, together with the uncertainty associated to the star-formation rate (orange region). Other spectra correspond to the case of mass-varying neutrinos (Tyrian purple), the case where both masses and mixing angles vary (green), for $z_s=0.05$ and ${\rm B}_s=5$, and the case where all neutrino masses are zero (cyan). We assume the normal ordering for the neutrino masses. The gray regions correspond to those where the expected number of background events is dominant.}\label{fig:Fvs_DUNE}
\end{figure}

Of course, a measurement of the DSNB in either Cherenkov or Liquid Argon detectors is not an easy task.
There are many sources of uncertainty and backgrounds that will impact the search for the DSNB. We do not address these in any detail here but instead, provide a simple example. 
We compute the number of events at a DUNE-like detector fixing ${\rm B}_s=5, z_s=0.05$, assuming an exposure of 400~kton-years, and considering as detection channel the process $\nu_e+\ ^{40}{\rm Ar}\to \ ^{40}{\rm K^*}+e^-$. As far as other characteristics of the DUNE-like detector, we repeat the assumptions we made in  Ref.~\cite{DeGouvea:2020ang}.
Fig.~\ref{fig:Fvs_DUNE} depicts event spectra as a function of the electron kinetic energy. The standard case is depicted in orange along with the uncertainties associated to our imperfect understanding of the SFR (orange region). We consider the case where only the neutrino masses vary (Tyrian purple) and the one where both the masses and mixing angles vary (green). The blue curve corresponds to the expected $\nu_e$ flux under the assumption that the neutrinos are massless.
The gray bands correspond to regions where background events are expected to be dominant.
If the neutrino masses ``turn on'' at a finite redshift, the DSNB $\nu_e$ flux is significantly smaller relative to standard expectations, as observed in the previous section.
For $E_{e^-}\gtrsim 30$~MeV, in the case where both masses and mixing angles vary (green), the flux is expected to lie slightly below the orange-shaded standard region.
Fig.~\ref{fig:Fvs_DUNE} reveals that a better understanding of systematic uncertainties is crucial to test the hypothesis that the neutrino oscillation parameters is $z$-dependent. As previously mentioned, a high-statistics measurement of the  DSNB antineutrino flux should play a decisive role in reducing uncertainties.

%
\section{Conclusions}\label{sec:conc}
\setcounter{equation}{0}
%
After more than two decades, the mystery surrounding the origin of the neutrino mass persists. A popular direction to pursue is to introduce new physics at high or very high energy scales but it is clear that very light new physics can also do the job. There is the possibility that the neutrinos are effectively massless at high redshifts and gain masses only recently, at very low redshifts, due to some exotic new physics operating at these scales. Such low-scale physics only affects the evolution of the Universe after photon decoupling and hence is completely compatible with observations of the CMB and other cosmic surveys. 

In this work, we propose that imprints of such low redshift neutrino-mass generation can be found on the diffuse supernova neutrino background (DSNB). The DSNB consists of neutrinos from all past supernovae (SNe), since the birth of star formation (redshifts around $5$). If neutrino masses are generated at relatively low redshifts, neutrino flavor-evolution through the SN is different from standard expectations. These effects can lead to significant changes to the flavor content of the neutrinos arriving at the Earth. Using a phenomenological parametrization for the redshift evolution of neutrino masses and mixing angles, we  computed the DSNB spectra at the Earth. We found that the DSNB $\nu_e$ spectral shape is sensitive to the epoch of neutrino mass generation: the peak can be, roughly, larger by up to a factor $1.5$, while the tail can be suppressed, leading to a more pinched spectrum. We also identified scenarios where, earlier in the history of the Universe, neutrino flavor propagation is completely non-adiabatic inside the SN. Finally, we simulated DSNB event spectra in a DUNE-like detector for different hypotheses concerning the time-dependency of the neutrino oscillation parameters and demonstrated that redshift-varying neutrino masses and mixing angles can lead to the suppression of the $\nu_e$ event spectrum. We find that there are circumstances under which effects due to time-dependent oscillation parameters are significant even if one includes uncertainties associated with our current understanding of the SFR, especially in the higher energy bins. The concurrent measurement, with enough statistics, of both electron neutrinos and antineutrinos from the DSNB, should allow one to make relatively robust claims about the constancy of neutrino masses.  

Measurements of the DSNB are possibly the only way to test scenarios where the mass-generation of neutrinos occurred only recently in the history of the Universe. The Super-Kamiokande experiment, doped with gadolinium, is expected to make a compelling discovery of the DSNB within this decade~\cite{Super-Kamiokande:2021the}. Future experiments like Hyper-Kamiokande (also doped with gadolinium) and DUNE are expected to collect a significant sample of DSNB events. On the astrophysical front, we expect the uncertainties on the SFR to go down in the coming decades. As a result, it is exciting to wonder whether a measurement of the DSNB can shed some light on the origin of the neutrino mass.

\section*{Acknowledgements}
We would like to thank Pedro Machado for illuminating discussions on mass-varying neutrinos. IMS, YFPG, and MS would like thank Northwestern University where part of the work was done. IMS and YFPG would like to thank the Fermilab theory group where this work started and for the lovely years well spent there. This work was supported in part by the US Department of Energy (DOE) grant \#de-sc0010143 and in part by the National Science Foundation under Grant Nos.~PHY-1630782 and PHY-1748958. 

\bibliographystyle{kpmod}
\bibliography{DSNB.bib}
\end{document}